\documentclass[aps,twocolumn,superscriptaddress,longbibliography]{revtex4-1}

\usepackage{amsmath}
\usepackage{amssymb}
\usepackage{amstext}
\usepackage{amsopn}
\usepackage{amsfonts}
\usepackage{amsxtra}
\usepackage[english]{babel}
\usepackage{graphicx}
\usepackage{bm}
\usepackage{multirow}
\usepackage{dcolumn}
\usepackage{color}
\usepackage{hyperref}
\usepackage{todonotes}
\usepackage{verbatim}
\def\LBCOx{La$_{2-x}$Ba$_{x}$CuO$_{4}$}
\def\LBCO{La$_{1.875}$Ba$_{0.125}$CuO$_{4}$}
\def\LBCOU{La$_{1.905}$Ba$_{0.095}$CuO$_{4}$}
\def\YBCO{YBa$_{2}$Cu$_{3}$O$_{6+\delta}$}

\def\mathbi#1{\ensuremath{\textbf{\em #1}}}
\def\Q{\mathbi{Q}}
\def\QCDW{\ensuremath{\mathbi{Q}_{\text{CDW}}}}
\def\TCDW{\ensuremath{T_{\text{CDW}}}}
\def\QSDW{\ensuremath{\mathbi{Q}_{\text{SDW}}}}

\begin{document}

\title{Incommensurate phonon anomaly and the nature of charge density waves in cuprates}

\author{H. Miao} \email{hmiao@bnl.gov}
\affiliation{Condensed Matter Physics and Materials Science Department, Brookhaven National Laboratory, Upton, New York 11973, USA}

\author{D. Ishikawa}
\affiliation{Materials Dynamics Laboratory, RIKEN SPring-8 Center, RIKEN, Sayo, Hyogo, Japan}
\affiliation{Research and Utilization Division, SPring-8/JASRI, Sayo, Hyogo, Japan}

\author{R. Heid}
\affiliation{Institut f\"{u}r Festk\"{o}rperphysik, Karlsruher Institut f\"{u}r Technologie, P.O. Box 3640, D-76021 Karlsruhe, Germany}

\author{M. Le Tacon}
\affiliation{Institut f\"{u}r Festk\"{o}rperphysik, Karlsruher Institut f\"{u}r Technologie, P.O. Box 3640, D-76021 Karlsruhe, Germany}

\author{G. Fabbris}
\author{D. Meyers}
\affiliation{Condensed Matter Physics and Materials Science Department, Brookhaven National Laboratory, Upton, New York 11973, USA}

\author{G. D. Gu}
\affiliation{Condensed Matter Physics and Materials Science Department, Brookhaven National Laboratory, Upton, New York 11973, USA}

\author{A. Q. R. Baron}
\affiliation{Materials Dynamics Laboratory, RIKEN SPring-8 Center, RIKEN, Sayo, Hyogo, Japan}

\author{M. P. M. Dean}\email[]{mdean@bnl.gov}
\affiliation{Condensed Matter Physics and Materials Science Department, Brookhaven National Laboratory, Upton, New York 11973, USA}

\date{\today}

\begin{abstract}

While charge density wave (CDW) instabilities are ubiquitous to superconducting cuprates, the different ordering wavevectors in various cuprate families have hampered a unified description of the CDW formation mechanism. Here we investigate the temperature dependence of the low energy phonons in the canonical CDW ordered cuprate \LBCO{}. We discover that the phonon softening wavevector associated with CDW correlations becomes temperature dependent in the high-temperature precursor phase and changes from a wavevector of 0.238 reciprocal space units (r.l.u.)\ below the ordering transition temperature up to~0.3 r.l.u.\ at 300~K. This high-temperature behavior shows that ``214''-type cuprates can host CDW correlations at a similar wavevector to previously reported CDW correlations in non-``214''-type cuprates such as \YBCO{}. This indicates that cuprate CDWs may arise from the same underlying instability despite their apparently different low temperature ordering wavevectors.
\end{abstract}

\maketitle

\section{Introduction\label{Introduction}}
Strongly correlated materials, such as the high temperature superconducting cuprates, exhibit remarkably rich phase diagrams, due in large part to strong coupling between their charge, spin and lattice degrees of freedom. Measuring how phonons are modified in the presence of CDWs can therefore provide compelling insights about the detailed CDW properties \cite{Pintschovius1991, McQueeney1999, Uchiyama2004,  Reznik2006, *Reznik2008, Graf2008, Astuto2008, Baron2008, Raichle2011,Reznik2012, Astuto2013, LeTacon2014}. Early work examining phonon dispersion in the cuprates tended to focus on Cu-O bond-stretching and bond-buckling phonon modes and identified a strong broadening and softening with cooling around $(0.25,0)$ in reciprocal lattice units (r.l.u.) \cite{Pintschovius1991, Reznik2006, *Reznik2008, Baron2008, Reznik2012, McQueeney1999, Uchiyama2004, Raichle2011, Graf2008}. These phonon anomalies have been widely discussed in terms of fluctuating stripes/CDWs that were originally predicted theoretically soon after the discovery of high $T_c$ superconductivity \cite{Zaanen1989, Machida1989, Poilblanc1989, Emery1990} and that are thought to freeze into an ordered state via coupling to octahedral tilts in the lattice \cite{Tranquada1995}. More recently, cuprate CDWs have gained renewed considerations due to the discovery of short ranged CDW correlations in essentially all $1/8$ doped cuprates \cite{Ghiringhelli2012, Chang2012, Comin2014, Neto2014, Thampy2014, Hashimoto2014, LeTacon2014, tabis2014, Miao2017} with a large phonon anomaly occurring on the transverse acoustic and optical phonon modes \cite{LeTacon2014}. While the presence of these correlations indicates that they may be an intrinsic property of superconducting cuprates, there are notable differences in the phenomenology \cite{Comin2016, Fradkin2015, Keimer2015}. Static CDW order in $1/8$ doped lanthanum-based ``214'' type cuprates emerges at low temperatures around $(0.25,0)$ with an incommensurability that appears locked to the spin density wave (SDW) correlations around $(0.375,0.5)$ (i.e.\ with an incommensurability of $1/8$ relative to the antiferromagnetic wavevector). For this reason CDWs in ``214'' type materials are often called ``stripes'' in analogy to phases in insulating nickelates \cite{Chen1993, Tranquada1995}. Stripe formation is thought to be driven primarily by frustrated spin interactions induced by doping a Mott insulator \cite{Zaanen1989, Machida1989, Poilblanc1989, Emery1990}. CDWs in non-``214'' cuprates appear at different wavevectors (e.g.\ $(0.3,0)$ for \YBCO{} (YBCO)). This wavevector shows no clear relationship with the location of the low energy spin excitations, which tend to be gapped. Non-``214'' cuprates also have smaller incommensurabilities at higher doping levels; whereas ``214'' systems tend to follow the Yamada relationship where the CDW wavevector increases with doping in concert with changes in the spin density wave (SDW) incommensurability \cite{Ghiringhelli2012, Fujita2012}. Such differences pose a significant challenge to a unified description of CDWs in all cuprates.

%
\begin{figure}[tb]
\includegraphics[width=8.5 cm]{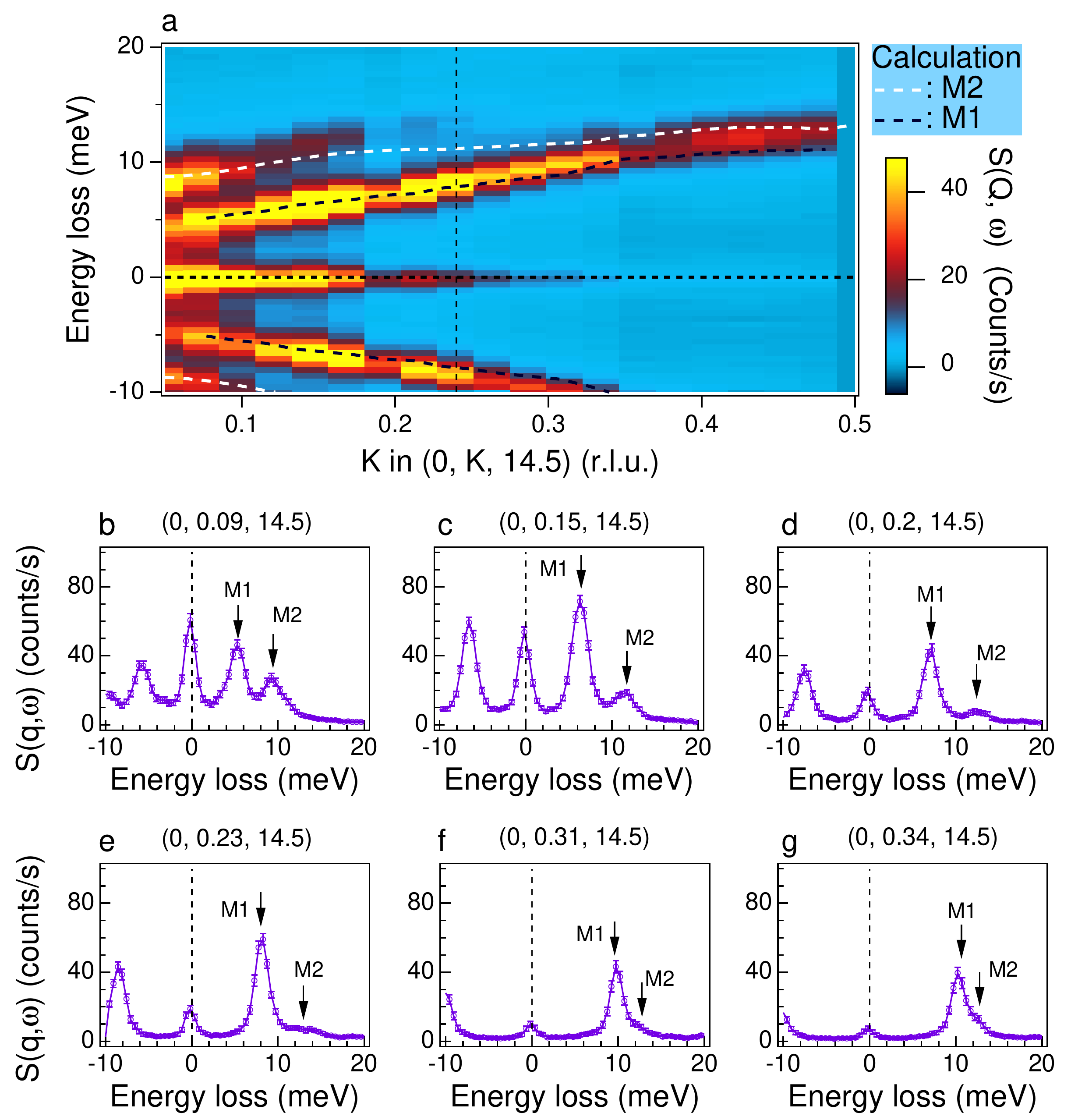}
\caption{Room temperature phonon dynamic structure factor $S(\Q,\omega)$ of \LBCO{}. (a) Colormap around $(0, K, 14.5)$ for $K=0\rightarrow0.5$~r.l.u. showing two modes labeled M1 and M2 that match theoretical predictions shown as dotted lines. (b) - (g) show representative IXS spectra at $K = 0.09$, $0.15$, $0.2$, $0.23$, $0.31$ and $0.34$~r.l.u., respectively.}
\label{Fig1}
\end{figure}

Here we report phonon dispersion measurements of archetypal CDW ordered \LBCOx{}, $x=0.125$ and $x=0.095$ \cite{Moodenbaugh1988, Fujita2004}, in which we focus on the low energy phonon modes associated with $c$-axis displacements complementing previously studied high energy optical phonons \cite{Reznik2006, *Reznik2008, Reznik2012, McQueeney1999, Uchiyama2004, Raichle2011, Graf2008}. The comparatively high spectral intensity of the low-energy  modes, alongside progress in inelastic x-ray scattering (IXS) instrumentation \cite{Baron2010,Baron2014}, allows for measurements with excellent wavevector, \Q{}, and energy, $\omega$, resolution and an extensive temperature dependence. For $x=0.125$ we find that CDW ordering at 55~K induces a small, narrow phonon anomaly near $(0.24,0)$, whereas a short-ranged CDW correlation is established at much higher temperatures giving rise to a broad phonon anomaly with a temperature-dependent wavevector that evolves from $0.24$~r.l.u.\ below 55~K to $0.30$~r.l.u.\ at 300~K. This result has significant consequences. First, it shows that both the ``214" and non-``214" cuprate can host CDW correlations around  0.3~r.l.u.\ Second, combined with previous results \cite{Fujita2004}, our data implies that CDW and SDW are unlocked at high-temperatures. This behavior is captured by Landau theory modeling of CDW order where a temperature dependent wavevector arises due to CDW and SDW correlations, in isolation, minimizing their energy at different wavevectors at high temperature \cite{Zachar1998}. Upon cooling, coupling between the CDW and SDW becomes important and locks the CDW and SDW incommensurabilities together at low temperatures \cite{Zachar1998}. In summary, our measurement of a temperature-dependent CDW-related phonon softening thus indicates that cuprate CDWs can arise from the same underlying instability despite their apparently different ordering wavevectors.

\section{Two anomalies in the phonons \label{anomalies}}
An IXS measurement of the dynamic structure factor, $S(\Q,\omega)$, of \LBCO{} along with the phonon mode dispersion predicted using Density Functional Perturbation Theory (DFPT) (App.~I-II and Supp.~I) are shown in Figure~\ref{Fig1}a. Representative IXS raw data are shown in Fig.~\ref{Fig1}b-g. The spectra are dominated by two modes, referred to here as M1 and M2, in good agreement with the the DFPT calculations. From the calculations, we ascertain that these modes are primarily associated with $z$ and $y$ direction motions of the heavier La and Cu atoms. M1 has primarily transverse acoustic character and M2 has mixed longitudinal acoustic and longitudinal optical character. These modes connect to the transverse and longitudinal branches of the $[0,0,L]$ direction, respectively (Supp.~I). 
%
\begin{figure}
\includegraphics[width=8.5cm]{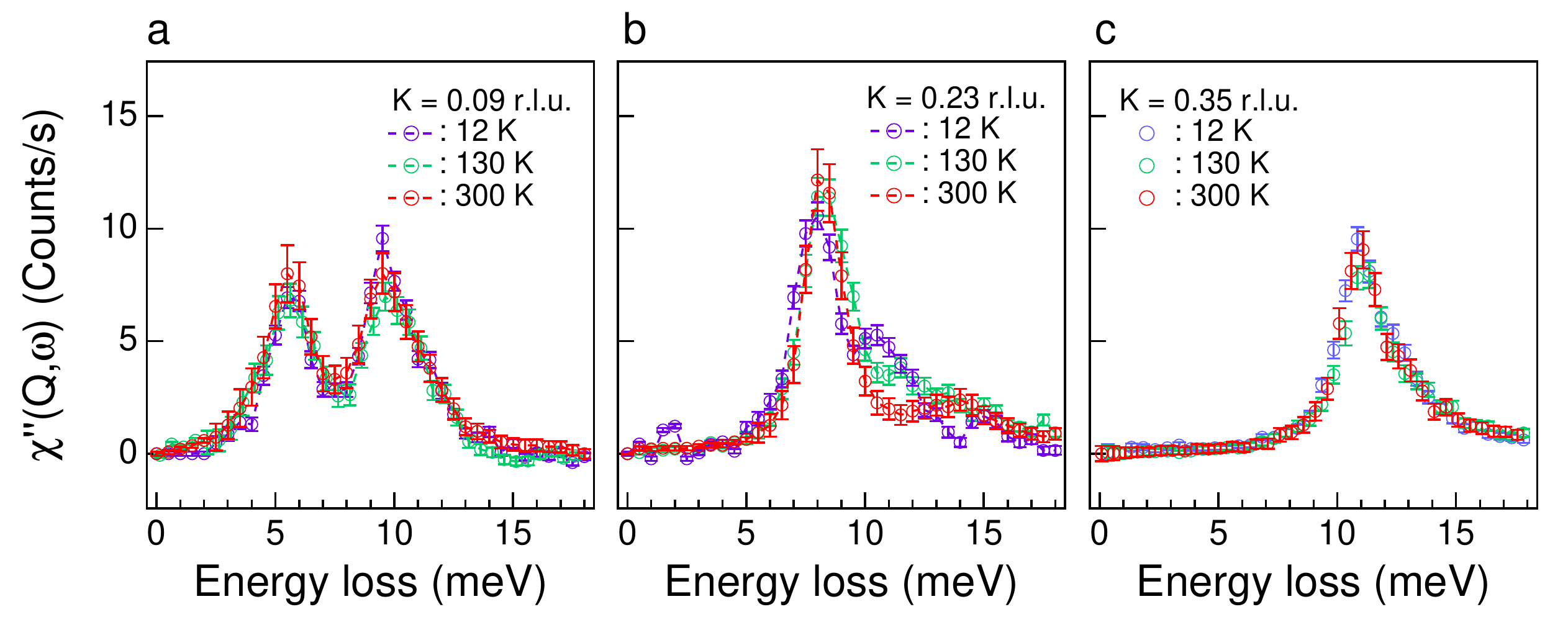}
\caption{Temperature dependence of the \LBCO{} phonon spectra. (a)-(c) Elastic line subtracted and Bose-factor corrected IXS spectra at $K = 0.09$, $0.23$ and $0.35$~r.l.u.\, respectively. Purple, green and red circles are correspond to data at 12, 130 and 300~K, respectively. Strong temperature dependence is seen at $0.23$~r.l.u., whereas the $K = 0.09$ and $0.35$~r.l.u. show no obvious temperature dependence.
\label{Fig2}}
\end{figure}

In order to examine the temperature dependence of the signal, we plot the imaginary part of the phonon dynamic susceptibility $\chi^{\prime\prime}(\Q,\omega)$ in Fig.~\ref{Fig2}, which is derived by subtracting the elastic line and correcting for the thermal Bose-factor as
\begin{equation}
\chi^{\prime\prime}(\Q,\omega)=S(\Q,\omega)(1-e^{-\omega/k_{B}T}).
\end{equation}
Panels a-c show spectra at $K = 0.09$, $0.23$ and $0.31$~r.l.u., respectively. A strong softening of M2 with decreasing temperature is seen at \QCDW{}, which is reduced or absent at other \Q{}.
%
%
\begin{figure*}
\includegraphics[width=1.0\textwidth]{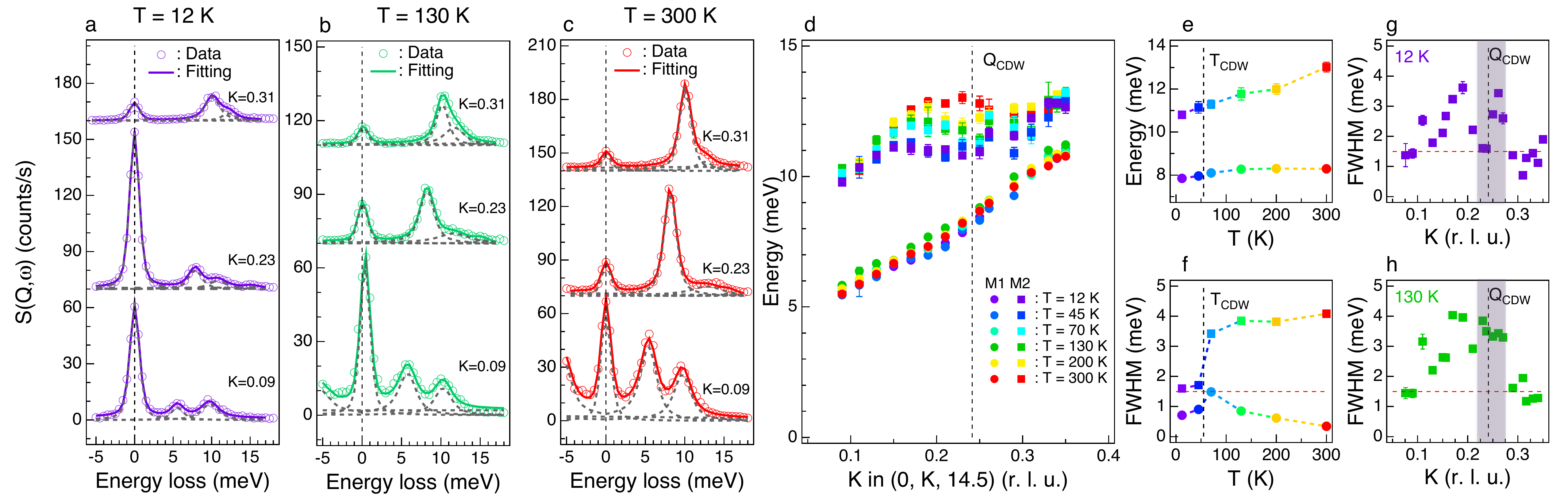}
\caption{Phonon lineshape analysis in \LBCO{}. (a)-(c) Example fits of spectra at $K=0.09$, 0.23 and 0.31~r.l.u. at temperatures of 12, 130 and 300~K, respectively. The individual phonon and elastic components are shown at dotted gray lines. Spectra at $K=0.23$ and 0.31~r.l.u.\ are vertically shifted for clarity. (d)-(h) Evolution of fitting parameters. Modes M1/M2 are shown as circles/squares and different colors denote different temperatures. (d) The energy dispersion along $(0, K, 14.5)$ through \QCDW{} (vertical dotted line). (e) \& (f) Temperature dependent peak energy and width, respectively, at $K=0.23$~r.l.u.\ (\QCDW{}). The vertical line marks \TCDW{}. (g) \& (h) show the \Q{}-dependent peak widths at 12 and 130~K, respectively. The shaded gray region at \QCDW{} shows the resolution limited width of the CDW Bragg peak at 12~K. The horizontal red line denotes the experimental energy resolution of 1.5~meV FWHM. Note that width values do not include the experimental energy resolution, as this is separately accounted for by the convolution in Supp.~Eq.~1. 
\label{Fig3}}
\end{figure*}

To extract the momentum and temperature dependent phonon properties, we fitted all measured IXS spectra along $(0, K, 14.5)$. Phonon modes are represented by standard damped harmonic oscillator functions \cite{Fak1997} where $A_{i}$, $\omega_{i}$ and $2\gamma_{i}$ are the intensity, energy and full width at half maximum (FWHM) of phonon peak $i$, that are weighted by the Bose-factor and convoluted with the 1.5~meV energy resolution function \footnote{See section II of the Supplementary Materials for a discussion of the possible role of phonon splitting.}. An intensity-scaled resolution function is used to represent the elastic line (Supp.~I). We note that an enhanced elastic line seen at low temperature around the CDW ordering wavevector is energy-resolution-limited, consistent with static CDW order \cite{Chen2016}. Representative fitting results are shown in Fig.~\ref{Fig3}a-c and the parameter evolution is plotted in Fig.~\ref{Fig3}d-h. M2 shows a strong softening over a broad \Q{} range that is already visible at 200~K [Fig.~\ref{Fig3}d\&e]. Based on the \Q{}-dependence of the effect, we assign it to phonons coupling to CDW correlations. 

Figure~\ref{Fig3}f-h shows that M2 also tends to be strongly broadened by almost 4~meV (FWHM) in a similar \Q{} range to the softening in contrast to M1 that remains resolution limited at all temperatures. This broadening effect on M2 is largely independent of temperature from 300 to 70~K, until a sharp reduction in the linewidth occurs upon cooling from 70 to 45~K. This reduction in linewidth is localized to a small \Q{}-window around \QCDW{}. 

In view of the fact that M2 couples more strongly to the CDW than M1, we re-examined the displacements associated with these modes using DFPT (Supp.~I). Although at half integer $L$, neither M1 nor M2 can be assigned to a purely longitudinal or transverse mode, we find that M2 has a larger component of $c$-axis displacements compared to M1. We therefore suggest that $c$-axis atomic displacements couple most strongly to the CDW. Due to the quasi-two-dimensional electronic structure of the cuprates, $c$-axis displacements are expected to be much more weakly screened out facilitating this stronger coupling. This conclusion is also consistent with studies that have associated cuprate CDWs with out-of-plane oxygen displacements \cite{Tranquada1995, Forgan2015, Bozin2016}. Indeed, to reduce the strong inter-layer Coulomb repulsion, the phase of CDW in neighbored unit cells is shifted by $\pi$ and yields CDW wavevector that is broadly peaked at half-integer $L$. 

\section{Relationship between phonons and CDW properties \label{Phonon-CDW}}
How do we relate these anomalies in the lattice dynamics to the underlying CDW correlations in the cuprates? Upon cooling, \LBCO{} goes through a transition at $\TCDW=55$~K\textbf{} below which long-ranged ($\sim$210~\AA) static CDW order emerges \cite{Fujita2004, Hucker2011, Wilkins2011, DeanLBCO2013, Miao2017, Chen2016, Thampy2017}. The phonon narrowing effect outlined in Sec.~\ref{anomalies} appears upon cooling below this transition [Fig.~\ref{Fig3}f] and its width in \Q{} is similar to that of the CDW Bragg peak [Fig.~\ref{Fig3}g]. We consequently associate the narrowing with the onset of the ordered CDW phase. 

\begin{figure}
\includegraphics[width=5.2 cm]{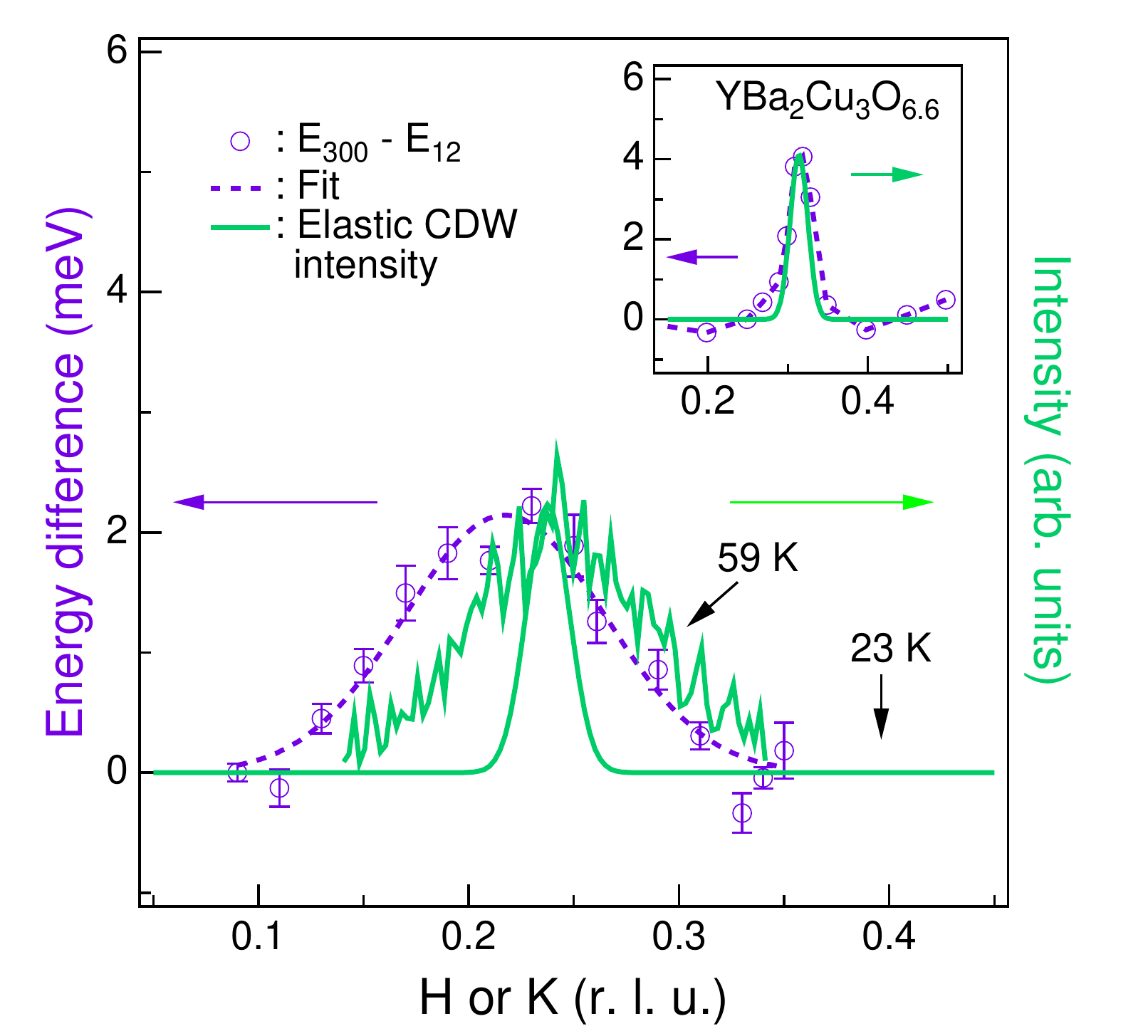}
\caption{Comparison between phonon softening and diffraction. Purple circles plot CDW-related phonon softening. The main panel plots the energy difference between 300 and 12~K for \LBCO{}, whereas insets show softening between 300 and 5~K for \YBCO{}. Dashed purple lines are guides to the eyes. These are compared to diffraction data in green. For \YBCO{} \cite{LeTacon2014} the phonon softening mirrors the CDW Bragg peak. In \LBCO{} the phonon softening does not match the ordered CDW Bragg peak taken at 23~K; rather it matches precursor CDW measured at 59~K \cite{Miao2017}.
\label{Fig4}}
\end{figure}

Phonon softening and broadening is, however, already apparent at temperatures well above $\TCDW$. This has led previous \LBCO{} studies to speculate that the high temperature phonon softening is associated with precursor CDW correlations, i.e.\ intrinsic fluctuating correlations that condense into the ordered state upon cooling \cite{Reznik2006, *Reznik2008, Astuto2008, Reznik2012}. The recent diffraction measurement of diffuse quasielastic scattering associated with precursor correlations in \LBCO{} \cite{Miao2017} allows us to directly test the relationship between diffraction measurements of the CDW order parameter and phonon softening. Figure~\ref{Fig4} overlaps the phonon softening and diffraction data seen in \LBCO{} and compares it to that observed in cuprate \YBCO{} \cite{LeTacon2014}. As seen in the inset, the CDW induced phonon softening closely matches the CDW Bragg peak in \YBCO{}. In \LBCO{}, however, the phonon softening is an order of magnitude broader in momentum space than the width of the CDW Bragg peak at 23~K, suggesting it is not directly related to the CDW ordering. Instead, this broad phonon anomaly matches the width of the quasi-elastic precursor-CDW intensity observed by resonant inelastic x-ray scattering (RIXS) above \TCDW{} \cite{Miao2017}, explicitly confirming its association with the precursor-CDW as distinct from the ordered phase. This is our first main observation.

%
%

Our results paint a different picture to conventional CDW systems \cite{Hoesch2009, Chatterjee2015, Weber2011}. These systems tend to exhibit phonons that soften to zero energy at the CDW ordering transition, and such softening occurs in a relatively narrow temperature window around the transition. In some cases these effects are reasonably well understood. For example, DFPT including strong electron-phonon coupling and phonon anharmonicity reproduces the phonon anomaly in NbSe$_2$, including its broad momentum dependence, which has similar momentum-width to that observed here \cite{Weber2011}. One difference, however, is that in LBCO we observe only partial phonon softening that occurs continuously from 300 to 12~K. Combined with the fact that the phonon anomaly matches the width of the quasi-elastic precursor correlations, this result is more consistent with the idea that the anomaly reflects phonons coupling to the CDW correlations, rather than electron-phonon coupling driving CDW formation \cite{Zhu2015}.  In \LBCO{}, the continuous softening, with only a very small phonon narrowing through the transition may suggest a dominant role of the precursor-CDW and that only a small fraction of the correlations are condensed below the transition temperature. Cuprates are known to host appreciable disorder \cite{Alloul2009, Haskel2000}, and, as theoretically modelled \cite{Nie2017}, disorder can suppress the formation of long-range ordered phases from short range precursor correlations and may have an important role here.

%
\begin{figure}[tb]
\includegraphics[width=5.2 cm]{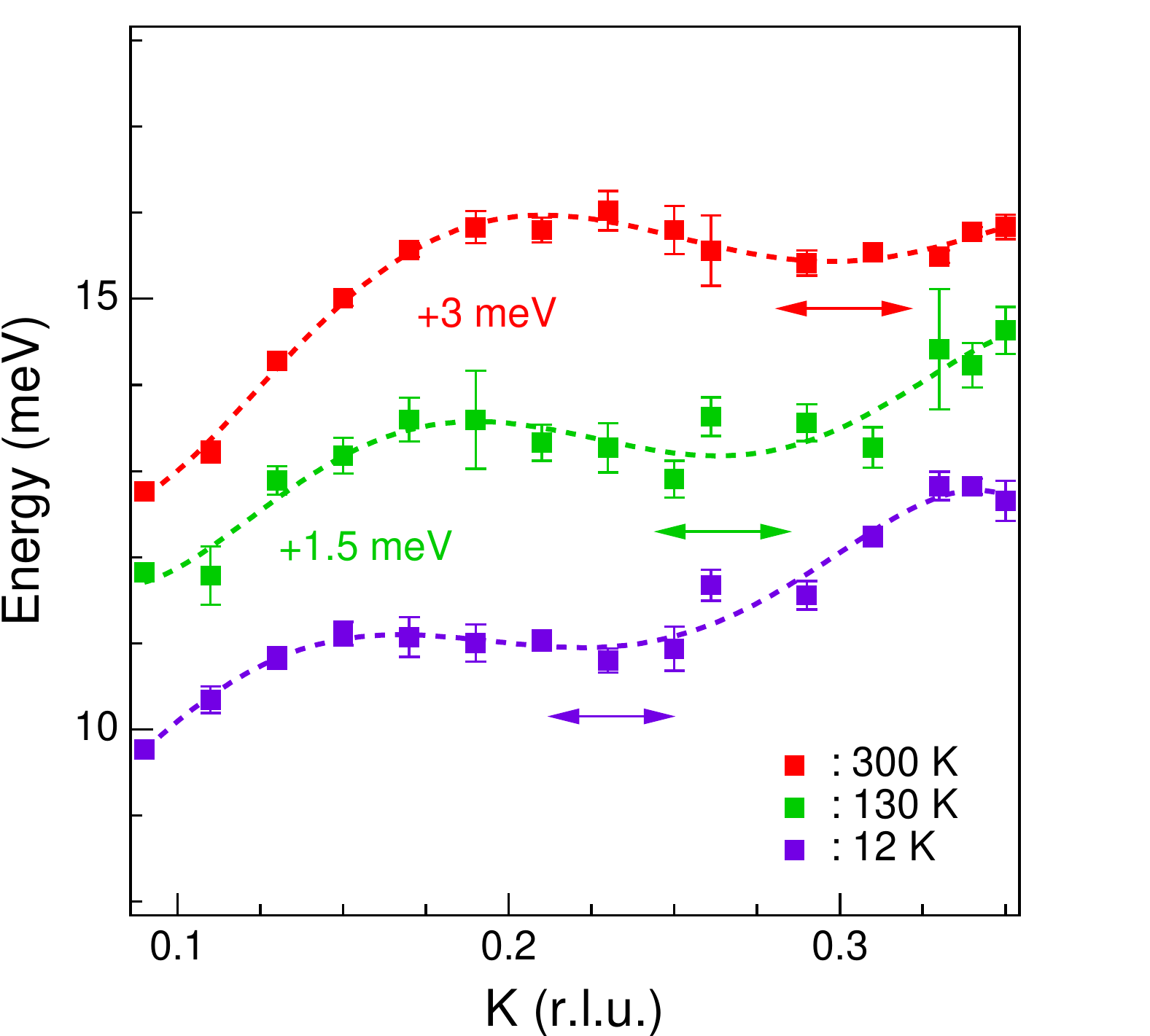}
\caption{Temperature dependent phonon incommensurability in \LBCO{}. Extracted phonon dispersions of M2 at 12, 130 and 300~K are shown as purple, green and red squares. Dashed lines are guides to the eyes and arrows indicate the \Q{} associated with the CDW related phonon softening. Dispersions at 130 and 300~K are offset by 1.5 and 3~meV, respectively for clarity.  
}
\label{Fig5}
\end{figure}


\section{Universal features of CDWs in cuprates \label{Unify}}
A notable difference between \LBCO{} and other cuprates is the wavevector of the CDW correlations. ``214'' type cuprates have wavevectors close to $0.25$~r.l.u. whereas other $1/8$ doped cuprates, tend to have higher wavevectors such as 0.3~r.l.u.\ for \YBCO{} \cite{Comin2016}. This motivated proposals that non-``214'' type cuprates should be understood using different paradigms than those in ``214'' systems \cite{Ghiringhelli2012,Comin2014,Wang2014,Comin2016,Liu2016,Neto2016}. In Fig.~\ref{Fig5} we examined the phonon dispersion and the CDW-related phonon softening in more detail. The maximum phonon softening, which is close to the CDW ordering wavevector at low temperatures, moves with increasing temperature, reaching around 0.3~r.l.u.\ at room temperature. Importantly, this unusual phonon softening is consistent with recent RIXS measurements of CDW-related quasi-elastic scattering in the same material \cite{Miao2017}, proving that the change of phonon softening is driven by the temperature dependent CDW incommensurability. This is our second main observation. The strong precursor CDW correlation and its unusual temperature-dependence can be compared to similar acoustic phonon dispersion measurements in \YBCO{} \cite{LeTacon2014}. Phonons in both \LBCO{} and \YBCO{} soften by a similar order of magnitude: 19\% and 26\% and both effects are already evident near 0.3~r.l.u. at the first temperature measured below the initial room temperature measurement 200 and 150~K, respectively. The similar CDW-associated phonon anomalies are consistent with the similar CDW phase diagram in these two cuprate families and indicates, albeit circumstantially, that both CDWs may share a common origin.

\section{Discussion of strong correlation induced CDW \label{Tdep_dis}}
In cuprates, the origin of the CDW is under debate between reciprocal space pictures such as Fermi-surface nesting and real space pictures with strong magnetic and Coulomb interactions. In the first class of scenarios, the most important property of the ground state is the wavevector which nests regions with a high density of states on the Fermi-surface such as the anti-nodal Fermi-surface \cite{Ghiringhelli2012, Neto2016} or the end-points of the Fermi-arcs \cite{Comin2014}. Here, the temperature dependence of the incommensurability (Sec.~\ref{Unify}) as well as the broadness of the phonon softening in momentum space (Sec.~\ref{anomalies}) do not match theoretical expectations based on Fermi surface. These scenarios would predict that the CDW wavevector should either decrease at higher temperatures \cite{Valla2006} or be temperature independent, neither of which is observed here. Instead, real space pictures are likely more important. These include those based on the competition between minimizing the number of broken magnetic bonds and kinetic energy and Coulomb repulsion between the doped holes \cite{Zaanen1994,Poilblanc1989, Emery1990, Vojta2009, Wagner2015, Huang2016}. In such scenarios the ordering wavevector arises from a balance between different ordering tendencies and is not a crucial defining parameter of the mechanism. These scenarios are more compatible with our observed variation of the CDW wavevector with temperature, which implies that the CDW wavevector is not solely defined by doping. Instead it changes on a thermal energy scale of order 10~meV. It should be noted that a similar conclusion has been put forward based on STM studies of Bi$_{2}$Sr$_{2}$CaCu$_{2}$O$_{8+x}$, where the authors found that the local CDW wavevector is doping independent and the change of the wavevector is due to ``phase slip" in the CDW domain boundaries hence not a fundamental property of the CDW \cite{Mesaros2016}.

\section{Spin-charge locking in L\lowercase{a}$_{2-x}$B\lowercase{a}$_{x}$C\lowercase{u}O$_4$ \label{Spin-charge}}
Since underdoped cuprates have a large magnetic energy scale and a relatively small electronic density of states at the Fermi level, much of the early work on CDW and SDW ordered phases in the cuprates assumed a dominant spin degree of freedom \cite{Zaanen1989,Machida1989, Poilblanc1989, Emery1990, Kato1990}. Indeed, the low-temperature CDW/SDW ordered phases in ``214'' materials have incommensurabilities that follow the ``Yamada'' rule in which  $\QCDW = (2\delta, 0)$ and $\QSDW  = (0.5-\delta, 0.5)$ with the incommensurability $\delta \approx x$. Here we see that the CDW incommensurability increases with temperature. From previous inelastic neutron scattering measurements of \LBCO{}, we know that the SDW incommensurability decreases with temperature \cite{Fujita2004}. Thus the spin-charge locking is broken at high temperature showing that \LBCOx{} can host a spin-charge unlocked phase very similar to non-``214'' cuprates which also host CDWs with no related low energy SDW correlations (the spin excitation spectrum in \YBCO{} is, in fact, gapped). We further tested this by measuring \LBCOU{} (plotted in Supp.~IV), finding that the maximum softening also occurs at larger wavevectors than the CDW ordering wavevector of 0.21 r.l.u.\ again unlocked from the SDW wavevector \cite{Hucker2013}. All these results support the case for a universal CDW mechanism in the cuprates. Temperature-dependent incommensurabilities were actually predicted by Ginzberg-Landau-Wilson modeling of coupled CDW and SDW order parameters \cite{Zachar1998, Zachar2000, Nie2017}. In such an approach, isolated CDW and SDW orders are assumed to minimize their energy at unrelated wavevectors, as we observe at high temperatures. Upon cooling, however, a cubic coupling term means that the CDW and SDW can save energy by locking their incommensurabilities in the ground state. We finally note that the low temperature tetragonal (LTT) structural phase transition is also widely believed to have an important role for stabilizing CDW formation \cite{Tranquada1995}. We suggest that the LTT distortion is likely to have a minimal role for locking the CDW and SDW here because (i) changes in phonon wavevector are already apparent at temperatures well above the LTT transition \cite{Miao2017,Fujita2004} and (ii) the CDW in La$_{2-x}$Sr$_x$CuO$_4$ also occurs around wavevectors of 0.24 r.l.u. despite the fact that this system does not exhibit an LTT transition \cite{Thampy2014, Kimura2000}.

\section{Conclusions}
Our phonon dispersion measurements of \LBCO{} reveal two CDW-related phonon anomalies: a phonon softening associated with precursor CDW fluctuations as well as small phonon narrowing associated with CDW-ordering. The phonon softening associated with precursor correlations occur at a temperature-dependent wavevector changing from 0.238 to 0.3~r.l.u.\ as the system is heated from 12 to 300~K. These results indicate that the strong precursor-CDW reflects the intrinsic property of the CDW correlations that is similar in all underdoped cuprates, while the ordered phase reflects coupling between charge and spin correlations that locks the CDW and SDW together at low temperatures into a state that is particular to cuprates without a spin gap.

\begin{acknowledgements}
M.P.M.D.\ and H.M.\ acknowledge Philip Allen, Chris Homes, Jos{\'e} Lorenzana, Claudio Mazzoli, J\u{o}rg Schmalian and John Tranquada and for insightful discussions. IXS research by H.M.\ and M.P.M.D.\ is supported by the Center for Emergent Superconductivity, an Energy Frontier Research Center funded by the US Department of Energy (DOE), Office of Basic Energy Sciences. Work at Brookhaven
National Laboratory was supported by the U.S. Department of Energy, Office of Science, Office of Basic Energy Sciences, under Contract No. DE-SC00112704. The synchrotron radiation experiments were performed at BL43LXU in SPring-8 with the approval of RIKEN (Proposal No.\ 20160097 and 20170076). 
\end{acknowledgements}

\appendix
\section{Sample growth\label{Sample_growth}}
 La$_{2-x}$Ba$_{x}$CuO$_{4}$ single crystals were grown at Brookhaven National Laboratory using the floating zone method and mm sized samples were prepared from the resulting rod by cutting and cleaving. The barium concentration of the samples was confirmed by measuring the sample magnetization, which are known to be around 2~K for $x=0.125$ and 31~K for $x=0.095$ \cite{Moodenbaugh1988, Hucker2011}. The crystalinity of the samples was checked both prior to and during the IXS measurement and found to be of excellent quality with crystal mosaics of $\sim 0.02^{\circ}$. Numerous previous measurements on samples prepared in the same way further attest the sample quality \cite{Wilkins2011, Hucker2011, DeanLBCO2013, Thampy2013, Chen2016, Miao2017}. The wave vectors used in our manuscript are described using the high temperature tetragonal ($I4/mmm$) space group with $a = b = 3.78$~\AA{} and $c = 13.28$ and 13.24~\AA{} for $x=0.125$ and $x=0.095$, respectively.  
 
\section{Experimental setup\label{Experimental_setup}}
High precision phonon dispersion measurements of \LBCOx{} were made using the high-resolution IXS spectrometer installed at BL43LXU of SPring-8, Japan, which delivers world-leading x-ray flux for an experiment of this type \cite{Baron2010,Baron2014}. In order to maximize the effect on the phonons, we focus primarily on $x=0.125$ and measure the dispersion around $\QCDW = (0, 0.25, 14.5)$ where a strong CDW Bragg peak develops below $\TCDW=55$~K. A large $L$ geometry was chosen in order to enhance our sensitivity to $c$-axis displacements which are only weakly screened due to the quasi-two-dimensional electronic structure of the cuprates. All data were taken with an incident energy around 21.75~keV using the $(11, 11, 11)$ refection of silicon as a monochromator and analyzer. The instrumental resolution was found to be well described by a pseudo-Voight function
\begin{equation}
R(\omega) = (1-\alpha)\frac{A}{\sqrt{2\pi}\sigma}e^{-\frac{\omega^{2}}{2\sigma^{2}}}+\alpha\frac{A}{\pi}\frac{\gamma}{\omega^{2}+\gamma^{2}}
\label{resolution} 
\end{equation}
where $\alpha$ is confined between 0 and 1 and controls the relative fraction of the Lorentzian and Gaussian functions and $\sigma$ and $\gamma$ characterize the width of the two functions, respectively. The overall width of this function was $\sim1.5$~meV full width at half maximum (FWHM), which varies slightly depending on the analyzer. 

The sample was mounted with the \mathbi{H} and \mathbi{L} reciprocal lattice vectors close to the horizontal scattering plane. In this configuration a vertical column in the two-dimensional ($4\times6$) analyzer array closely traced our desired $(0, K, 14.5)$ trajectory in reciprocal space \cite{Baron2008, Baron2014}. Slits were used to define the momentum resolution of the measurement to be $(\delta\mathbi{H}, \delta\mathbi{K}, \delta\mathbi{L}) = (0.02, 0.02, 0.2)$~r.l.u. Where the relaxed value of $\delta\mathbi{L}$ was chosen in view of the known short correlation length of the CDW in this direction \cite{Hucker2011, Wilkins2011}. By examining spectra in the analyzers away from the main column of interest, we estimated the dispersion of M1 and M2 in the $L$ direction. From this, we calculated that finite $\delta\mathbi{L}=0.2$ \Q{}-resolution contributes only about 0.25~meV to the phonon linewidth, which is much smaller than the temperature-induced broadening effects discussed here. During the measurements, we interleaved multiple measurements such that neighboring points were separated by step of 0.02~r.l.u. in $K$. Spectra were collected for about 1 hour at each momentum transfer. The efficiency of each analyzer was first approximately determined by measuring plexiglass and further refined ($<$20\%) by scaling the values to ensure that the signal evolves smoothly as a function of \mathbi{Q}.

\bibliography{ref}
\end{document}